\documentclass[12pt]{article}
\usepackage{amsmath,verbatim,color,amssymb,epsfig,amsfonts,color,natbib}
\numberwithin{equation}{section}
\begin{document}
\newtheorem{lemma}{Lemma}
\newtheorem{theorem}{Theorem}
\newtheorem{remark}{Remark}
\begin{center}{\large \bf

Semiparametric Estimation of Average  Treatment Effect with Sieve Method
}
\end{center}

\begin{center}
{Jichang Yu\\
{\it School of Statistics and Mathematics, Zhongnan  University of   Economics and Law, Wuhan, Hubei 430073, China}\\

  Haibo Zhou and Jianwen Cai$^{\dag}$\\
{\it Department of Biostatistics, University of North Carolina at Chapel Hill, NC 27599, USA}\\

}
\end{center}

\let\thefootnote\relax\footnotetext{$^{\dag}$ Email: cai@bios.unc.edu}

\begin{abstract}
Correctly  identifying  treatment effects in observational studies is very difficult due to
the fact that the outcome model or the treatment assignment model must be correctly specified. Taking  advantages of semiparametric models in this article,
 we use  single-index models to establish the outcome model and the treatment assignment model, which can allow the link function to be unbounded and have
 unbounded support. The link function is regarded as a point in an
 infinitely dimensional function space, and we can estimate the link function
 and the index  parameter  simultaneously.
 The sieve method is used to approximate the link function and obtain the estimator of the average treatment effect by the simple linear regression. We establish the asymptotic properties of the proposed estimator.
 The finite-sample performance of the proposed estimator
 is evaluated through simulation studies and an empirical
example.
\end{abstract}

\vspace{0.5cm} \noindent{\bf Keywords}:  Hermite polynomials,  Observational studies, Propensity score, Sieve method, Single-index model

\noindent{\bf 2000 MR Subject Classification } 62G08, 62F12
\newpage
\section{Introduction}
 Correctly identifying  the effect of a policy or treatment is an important issue in   social and biomedical sciences \citep{imbens2015causal}. Randomised controlled trials (RCTs) are considered to be the ``gold standard'' for
treatment effects where the mechanism of the treatment assignment is random. However,
it is not always feasible to
conduct RCTs  due to political, financial   or ethical concerns. For example,
we can not randomly send people to college to evaluate the
 impact of college education on income;
we can not expose
people randomly to air pollution for evaluating the impact of PM2.5 on health.
Thus, much empirical work about treatment effects in economics and biomedical studies
need to rely on observational data, where the treatment assignment is not random or manipulated by investigators \citep{athey2017state}.

With rapid development of science and technology,   observational  data
containing fine-grained information about  markets and humans and their behavior are
becoming available, which  brings a great opportunity
to study   causal effects \citep{kunzel2019metalearners}. However,
it also brings a great challenge to correctly identify
treatment effects due to the lack of randomization in observational studies.
 Without randomization in the treatment allocation,
there usually
exist differences in the distribution of the baseline covariates between the treated and
control groups.
If these differences can not be reasonably adjusted, the estimator of treatment
effects will be seriously biased.
 The propensity score proposed by \cite{rosenbaum1983central} is commonly used to
balance the distribution of the baseline covariates in the sense that
 subjects with similar  propensity scores  will
   have  similar
distribution of the baseline covariates no matter which groups they are in.
 However, the propensity score is usually unknown in observational studies and  commonly  estimated by  the logistic  and Probit models in practice.

Since the work of \cite{rosenbaum1983central}, propensity score methods including the
  matching, stratification,
inverse probability weighting and covariate adjustment
 have been well studied by many authors (e.g., \citealp{lunceford2004stratification,austin2011introduction,hade2014bias,zou2016variance,
 hernan2020causal}).
  \cite{robins1994estimation} proposed an augmented inverse probability weighting estimator by
 combining the propensity score with the
  outcome regression model, which had been widely studied due to its double robustness.
 The properties of double robustness mean that
    the estimator will be  consistent if either the outcome regression model
  or the propensity score is correctly specified \citep{bang2005doubly}.

Correctly specifying the propensity score or the outcome model may not be
an easy task in practice,  furthermore if both of the above two models are misspecified, the doubly
robust estimator will be seriously biased \citep{kang2007demystifying}. \cite{tsiatis2007comment} also pointed out that the doubly robust estimator would be volatile when the  estimated propensity score is close to zero or one.  How to combine the outcome model and the propensity score to
reduce the bias caused by the misspecification of the two models is an important issue
in  causal inference.
Recently,
\cite{lee2018simple} proposed
a simple least squares estimator for treatment effects with the outcome regression
model being set to  a  polynomial function of the propensity score that  was estimated by
a Probit model, and thus the outcome regression model
 is a special case of the single index model.
\cite{wu2021semiparametric} proposed a semiparametric estimator
for the average causal effect using a propensity score-based spline with
the propensity score estimated
by a logistic model.

Parametric models heavily depend on their assumptions,
  such as the linearity assumption in the least
square regression, logistic regression and Probit regression.
However, there usually exists complex confounding structures
in the treatment assignment of observational data.
In order to take the full use of   advantages of semiparametric models,
 we use  single-index models to establish the outcome model and the treatment assignment model and allow the link function is unbounded and has
 unbounded support in this article. The link function is regarded as a point in an
 infinitely dimensional function space, which allows us to estimate the link function
 and the index  parameter  simultaneously. The link function and the index parameter
 can be derived from an optimization problem with constraints for the identification
 condition for the index parameter. Then, the estimator of the average treatment effect
 can be obtained by the simple linear regression.

The rest of the article is organized as follows.  In Section 2,  we introduce the
semiparametric models and statistical inference procedures.
In Section 3, we establish the asymptotic properties of the proposed estimator.
Simulation studies are conducted to evaluate the finite-sample performance of
the proposed estimator in Section 4.
A real dataset from  Pennsylvania is analyzed  to study the effect of
maternal smoking on birth weight  in
Section 5.
Some  conclusions and remarks are presented in Section 6.
The proofs are deferred to the Appendix.

\section{Semiparametric regression and inference procedures}
\subsection{Semiparametric regression model}
In order to simplify expressions, let $Y$ denote the outcome of interest, a binary variable $D$ denote the treatment
assignment ($D=1$ for  treatment, $D=0$ for control) and
$X=(X_1,\ldots,X_p)^{'}$ denote a $p$-dimensional vector of the pre-treatment covariates.
The observed data $\{(Y_i,D_i,X_i), i=1,\ldots, n\}$ are assumed to be
 independent copies of $(Y, D, X)$. We consider the following semiparametric regression model
 \begin{equation}\label{Semi_RM}
 Y=\alpha D +r(X)+\epsilon,
 \end{equation}
where $\alpha$ denotes the treatment effect, $r(X)$ is an unknown smoothed function and
$\epsilon$  is a mean-zero random error.
The  semiparametric model (\ref{Semi_RM}) had been widely studied
in the literature of econometrics and statistics (e.g., \citealp{engle1986semiparametric,robinson1988root,stock1991nonparametric}).    \cite{robinson1988root} proposed an innovative method for estimating the parameter $\alpha$ by rewriting model \eqref{Semi_RM} as follows:
 \begin{equation}\label{Robin_RM}
 Y-E[Y|X]=\alpha (D-E[D|X])+\epsilon.
 \end{equation}
The estimator of  $\alpha$ obtained via the simple linear regression model (\ref{Robin_RM})
with  $E[Y|X]$ and $E[D|X]$ estimated by the Nadaraya-Watson kernel method
had been proved to be root-$N$-consistent  by \cite{robinson1988root}.

Although Robinson's estimator is root-$N$-consistent, there exist several limitations
when applied to observational data to estimate the treatment effect. The performance of this semiparametric method heavily depends on the accuracy of the estimation of
$E[Y|X]$ and $E[D|X]$. The nonparametric  methods  are limited   due to the curse of dimensionality and the parametric model assumption may not capture the complicated relationship among the outcome, treatment assignment and baseline covariates in observational data. In order to avoid modeling $E[Y|X]$ based on individual component
of $X$,
\cite{lee2018simple} proposed to use the logistic regression  to model the propensity score $E[D|X]$ and then used
 the second or third order polynomials  of $X^{T}\widehat{\beta}$ to model $E[Y|X]$ with
 $\widehat{\beta}$ being the parametric estimator in the propensity score model.

In this article, we consider  single-index models for $E[Y|X]$ and $E[D|X]$, respectively,  which are given as follows:
\begin{eqnarray}\label{PS_single}
E[D|X]=\frac{\exp\left(g_{1,0}(X^{T}\beta_0)\right)}{1+\exp\left(g_{1,0}(X^{T}\beta_0)\right)}
\end{eqnarray}
and
\begin{eqnarray}\label{RG_single}
E[Y|X]=g_{2,0}(X^{T}\gamma_0),
\end{eqnarray}
where $\beta_0=(\beta_{0,1},\ldots, \beta_{0,p})^{T}$ and
$\gamma_0=(\gamma_{0,1},\ldots, \gamma_{0,p})^{T}$ satisfy
$\|\beta_0\|=1$ with $\beta_{0,1}\geq 0$ and
$\|\gamma_0\|=1$ with $\gamma_{0,1}\geq 0$ for identifiability
and the link functions $g_{1,0}(\omega)\in L^{2}(\mathbb{R},\pi(\omega))$ and  $g_{2,0}(\omega)\in L^{2}(\mathbb{R},\pi(\omega))$ with $\pi(\omega)=\exp(-\omega^2/2)$,
where $L^2(\mathbb{R},\pi(\omega))$ denotes the Hilbert space.
The Hilbert space $L^2(\mathbb{R},\pi(\omega))$ can cover many function classes, to
name a few, all polynomials, all power functions and all bounded functions on $\mathbb{R}$
(\citealp{chen2007large,dong2018additive,dong2019series}).
 Model (\ref{PS_single}) can cover the widely used logistic regression model
and Probit model as its  special cases. The method of \cite{lee2018simple} 
can be regarded  as a special case of our proposed method.

Single-index models have been well studied in the literature of statistics and econometrics. The non-parametric kernel method and  spline method are two commonly
used  methods to estimate  single-index models (\citealp{xia2006asymptotic,yu2002penalized,ma2015varying}). However, those methods  need to assume the
boundedness of the link function or its support.
Alternatively, sieve methods can provide  good approximations to unknown functions and are convenient to  calculate \citep{chen2007large}. In this article, we use the Hermite orthogonal  polynomials to approximate the link function in  single-index models, which do not need to assume the  link function is bounded and has bounded support.

\subsection{Inference procedures}
The Hermite polynomials form a complete orthogonal system in the Hilbert space
$L^{2}(\mathbb{R},\pi(\omega))$ with $\pi(\omega)=\exp(-\omega^2/2)$ and  its bases are
\begin{eqnarray*}
H_{m}(\omega)=(-1)^{m}\exp(\omega^2/2)\frac{d^{m}}{dw^{m}}\exp(-\omega^2/2),\, m=0,1,2,\ldots,
\end{eqnarray*}
which satisfy
$\int H_m(\omega)H_n(\omega)\pi(\omega)d\omega=m!\sqrt{2\pi}\delta_{mn}$ with $\delta_{mn}$ being the Kronecker delta.
Furthermore,  define
$h_m(\omega)=(\sqrt{2\pi}m!)^{-1/2}H_m(\omega)$, then
$\{h_m(\omega), m=0,1,2,\ldots\}$ becomes the standard orthogonal basis and satisfies
  $\int h_m(\omega)h_n(\omega)\pi(\omega)d\omega=\delta_{mn}$
in the Hilbert space
$L^{2}(\mathbb{R},\pi(\omega))$.
For any function $g(\omega)\in L^{2}(\mathbb{R},\pi(\omega))$, it has
an orthogonal series expansion in terms of $\{h_m(\omega), m=0,1,2,\ldots\}$ given as follows
\begin{eqnarray}\label{Ose_1}
g(\omega)=\sum\limits_{m=0}^{\infty}c_mh_m(\omega)\quad \mbox{with}\quad c_m=\int g(\omega)h_m(\omega)\pi(\omega)d\omega.
\end{eqnarray}
In the Hilbert space $L^{2}(\mathbb{R},\pi(\omega))$, the norm $\|\cdot\|_{L^2}$ is defined as
\begin{eqnarray*}
\|g\|_{L^2}=\left\{\int |g(\omega)|^{2}\pi(\omega)d\omega\right\}^{1/2},
\end{eqnarray*}
which is equal to $\sqrt{\sum\nolimits_{m=0}^{\infty}c_{m}^2}$ by Parseval's equality.
Therefore, the function $g(\omega)$ can be identified by the associated coefficients
$\{c_m, m=0,1, 2, \ldots\}$. For any truncation parameter $k\geq 1$,  the orthogonal series expansions can be split  into two parts:
\begin{eqnarray}\label{pt1}
g(\omega)=g_k(\omega)+\delta_k(\omega),
\end{eqnarray}
where $g_k(\omega)=\sum\nolimits_{m=0}^{k-1}c_mh_m(\omega)$ and $\delta_k(\omega)=
\sum\nolimits_{m=k}^{\infty}c_mh_m(\omega)$.
Under some regularity conditions, it is well known that $g_k(\omega)\rightarrow g(\omega)$ or
$\delta_k(\omega)\rightarrow 0$ as $k\rightarrow \infty$.
 The term $g_k(\omega)$ can be written as
$g_k(\omega)=\mathcal{H}(\omega)^{T}\mathcal{C}_k$ with $\mathcal{H}(\omega)=(h_0(\omega),
h_1(\omega),\ldots, h_{k-1}(\omega))^{T}$ and $\mathcal{C}_k=(c_0,c_1,\ldots, c_{k-1})^{T}$.

Due to  the expansion  (\ref{pt1}), the non-parametric function $g(\cdot)$ can be
parameterized with $\{c_m,m=0,1,\ldots\}$. To simply notations, we use $\theta$
for $\beta$ or $\gamma$ and $\theta_0$ for $\beta_0$ or $\gamma_0$, respectively.
The non-parametric function $g(\cdot)$
and the unknown parametric $\theta_0$
 can be reviewed as a point in an infinite-dimensional
Euclidean space, which is the two-fold Cartesian product space by $L^{2}(\mathbb{R},\pi(\omega))$ and $\mathbb{R}^{p}$ and is equipped with the norm
$\|\cdot\|$ being
\begin{eqnarray*}
\|(\theta, g)\|=\left(\sum\nolimits_{m=0}^{\infty}c_m^{2}+\|\theta\|^2\right)^{1/2}
\end{eqnarray*}
with $\|\theta\|=\sqrt{\theta_{1}^2+\cdots+\theta_{p}^2}$.
Suppose $\Theta$ is a compact set of $\mathbb{R}^{p}$ with $\theta_0$ being its interior point and $G$ is a subset of $L^{2}(\mathbb{R},\pi(\omega))$ with
$\sup\nolimits_{g\in G}\|g\|_{L^2}<M$ with $M$ being a large constant, $g_{1,0}\in G$ and $g_{2,0}\in G$. Define $G_k=G\cap \mbox{span}\{h_0(\omega),h_1(\omega),\ldots,h_{k-1}(\omega)\}$ with
$k$ being the truncation parameter.
Because
the treatment assignment  in model (\ref{PS_single}) is a binary variable, we define
\begin{eqnarray*}
L_{n}(\beta,g)=-n^{-1}\sum\limits_{i=1}^{n}\left\{D_i(\mathcal{H}(X_{i}^{T}\beta)^{T}
\mathcal{C}_k)-log\left(1+\exp(\mathcal{H}(X_{i}^{T}\beta)^{T}
\mathcal{C}_k)\right)\right\}.
\end{eqnarray*}
Because the outcome in model (\ref{RG_single})  is a continuous variable,
we define
\begin{eqnarray*}
L_{n}(\gamma,g)=n^{-1}\sum\limits_{i=1}^{n}\left[y_i-\mathcal{H}(X_{i}^{T}\gamma)^{T}
\mathcal{C}_k\right]^{2}.
\end{eqnarray*}
Then,
the estimators of $\theta_0$ and $\mathcal{C}_{0,k}$ can be obtained by solving the following constrained minimization problem
\begin{eqnarray}\label{est1}
(\widehat{\theta},\widehat{\mathcal{C}}_k)=\arg\min\limits_{\Omega_k,\lambda}
W_{n}(\theta,g)=\arg\min\limits_{\Omega_k,\lambda}\left[L_{n}(\theta,g)+
\lambda(\|\theta\|^{2}-1)\right],
\end{eqnarray}
where $\Omega_k=\{(\theta,\mathcal{C}_k): \theta\in \Theta, \|\mathcal{C}_k\|_2\leq M\}\subseteq \mathbb{R}^{d+k}$ with $\mathcal{C}_k=(c_0,\ldots, c_{k-1})^{T}$.
 The estimator for the average treatment effect
is given as follows
\begin{eqnarray}\label{est2}
\widehat{\alpha}_{n}=\frac{\sum\limits_{i=1}^n\left[y_i-\widehat{g}_{2,k}(X_{i}^{T}\widehat{\gamma})\right]
\left[D_i-\frac{\exp(\widehat{g}_{1,k}(X_{i}^{T}\widehat{\beta}))}
{1+\exp(\widehat{g}_{1,k}(X_{i}^{T}\widehat{\beta}))}\right]}
{\sum\limits_{i=1}^n\left[D_i-\frac{\exp(\widehat{g}_{1,k}(X_{i}^{T}\widehat{\beta}))}
{1+\exp(\widehat{g}_{1,k}(X_{i}^{T}\widehat{\beta}))}\right]^2},
\end{eqnarray}
where $\widehat{g}_{1, k}(X^{T}\widehat{\beta})=
\widehat{\mathcal{C}}_{1, k}^{T}\mathcal{H}(X^{T}\widehat{\beta})$
and $\widehat{g}_{2, k}(X^{T}\widehat{\gamma})=
\widehat{\mathcal{C}}_{2, k}^{T}\mathcal{H}(X^{T}\widehat{\gamma})$.

\section{Asymptotic properties}
In order to establish the asymptotic properties of the proposed estimators, we assume the following conditions hold.
\begin{itemize}
\item[C1]$\Theta \subset \mathbb{R}^{p} $ is a convex and compact set and   $\beta_0$ and $\gamma_0$ are two interior points of $\Theta$.  $G\subset L^{2}(\mathbb{R},\pi(\omega))$ with
$\sup\limits_{g\in G}\|g\|_{L^2}<M$ with $g_{1,0}\in G$, $g_{2,0}\in G$ and $M$ being a large constant.
\item[C2] Define $\Sigma_1(\theta_0,\mathcal{C}_{0,k})=E\left[(\mathcal{\dot{H}}
(X^{T}\theta_0)^{T}\mathcal{C}_{0,k})^{2}XX^{T}\right]$ and $\Sigma_2(\theta_0)=E\left[\mathcal{H}
(X^{T}\theta_0)\mathcal{H}(X^{T}\theta_0)^{T}\right]$,
we have $\varepsilon_1\leq\lambda_{min}(\Sigma_1(\theta_0,\mathcal{C}_{0,k}))\leq
\lambda_{max}(\Sigma_1(\theta_0,\mathcal{C}_{0,k}))\leq M_1$ and
$\varepsilon_2\leq\lambda_{min}(\Sigma_2(\theta_0))\leq
\lambda_{max}(\Sigma_2(\theta_0))\leq M_2$,
where $\varepsilon_1$, $\varepsilon_2$, $M_1$ and $M_2$ are positive constants,
 $\lambda_{max}(\Sigma)$ and  $\lambda_{min}(\Sigma)$ denote the maximum and
 minimum eigenvalues of matrix $\Sigma$.

\item[C3] The link function $g_{0}(u)$ is $m$-order differentiable on $\mathbb{R}$
and $g_{0}^{(j)}(u)\in L^{2}(\mathbb{R},\pi(u))$, $j=0,1,\ldots, m$, for some positive
integer $m$.
\item[C4] The truncation parameter $k$ is divergent with $n$ such that
$kn^{-1}\rightarrow 0$ and $nk^{-m}\rightarrow 0$ as $n \rightarrow \infty$,
where $m$ is defined in Condition C3.

\item[C5] $E[\epsilon|D,X]=0$ and $E[\epsilon^{2}|D,X]=\sigma^{2}$, where $\sigma^{2}$
is a positive constant.

\item[C6] $\sup\nolimits_{\{(\theta, u)\in \Theta\times \mathbb{R}\}}\exp(u^2/2)f_{\theta}(u)\leq M$, where $f_{\theta}(u)$
    is the probability density function of $u=X^{T}\theta$ and
    $\sup\nolimits_{\{(\theta, u)\in \Theta\times \mathbb{R}\}} E\|XX^{T}\{g^{(1)}(x^{T}\theta)\}\|\leq M$.
\end{itemize}
\begin{remark}
Condition C1 is commonly needed for extremum estimation \citep{chen2007large}.
Conditions C2 is commonly used in the literature of single-index models \citep{yu2002penalized}.
Condition C3 imposes some smoothness on the link function to ensure the negligibility
of the truncation residuals.
Condition C4 ensures the truncation
residuals can be smoothed out when we establish the asymptotic normality.
Condition C5 shows  the error is exogenous and homogeneous, which can be
extended to heteroscedastic error.
 Condition C6 excludes heavy-tailed distributions due to the fact the link function may be unbounded.

\end{remark}

\begin{theorem}\label{thm1}
Under Conditions C1-C4, we have
$\|(\widehat{\beta},\widehat{g}_{1,k})-(\beta_0,g_{1,0})\|\rightarrow_{p} 0$
as $n$ goes to infinity.
\end{theorem}
We defer the proof of Theorem \ref{thm1} to the supplemental materials.

\begin{theorem}\label{thm2}
Under Conditions C1-C6, we have
$\|(\widehat{\gamma},\widehat{g}_{2,k})-(\gamma_0,g_{2,0})\|\rightarrow_{p} 0$
as $n$ goes to infinity.
\end{theorem}
 The proof of Theorem \ref{thm2} is given in  the supplemental materials.

\begin{theorem}\label{thm3}
Under Conditions C1-C6,  $\sqrt{n}(\widehat{\alpha}_n-\alpha)$
converges in distribution to a zero-mean normal distribution  with variance
$\sigma^2\Psi^{-1}$,
where $\Psi=E[D-E[D|X]]^2$ and $\sigma^2$ is defined in Condition C5.
\end{theorem}
 The proof of Theorem \ref{thm3} is presented in  the supplemental materials.

\begin{remark}
Theorem \ref{thm3} shows the estimation errors of $\widehat{E}[D|X]-E[D|X]$
 and $\widehat{E}[Y|X]-E[Y|X]$ can be ignored when we establish the asymptotic properties of the proposed estimator for treatment effects, which  coincides with
  the conclusion of \citep{robinson1988root}.
\end{remark}

\section{Simulation study}
In this section, we conduct  simulation studies to study the finite-sample
performance of the proposed method.
We consider three scenarios to mimic the real world for casual effects  in observational studies.

{\it Simulation I:} \quad
The response is generated by the
 linear  regression model
\begin{eqnarray*}\label{simu_my1}
Y=\alpha D+\gamma_1X_1+\gamma_2X_2+\gamma_3X_3+\epsilon,
\end{eqnarray*}
where $D$ denotes the treatment assignment,  $X_1$, $X_2$ and $X_3$ follow the standard normal distribution, the error term
$\epsilon$  follows the standard normal distribution, $(\gamma_1,\gamma_2,\gamma_3)=(0.4,0,0.917)$ and $\alpha$ is equal to $0$ or $0.5$.
 The treatment assignment model is
\begin{eqnarray*}\label{simu_mt1}
P(D=1|X_1,X_2,X_3)=\frac{\exp(\beta_1X_1+\beta_2X_2+\beta_3X_3)}{1+
\exp(\beta_1X_1+\beta_2X_2+\beta_3X_3)},
\end{eqnarray*}
where $(\beta_1,\beta_2,\beta_3)=(0.8,0,-0.6)$.
It is obvious that the response model
and the treatment assignment model are  simple single-index models.

{\it Simulation II:} \quad
We consider the situation where the response model
and the treatment assignment model are more  complex single-index models.
The response is generated by the following partially single-index model:
\begin{eqnarray*}\label{simu_my2}
Y=\alpha D+\exp(\gamma_1X_1+\gamma_2X_2+\gamma_3X_3)+\epsilon.
\end{eqnarray*}
The treatment assignment model is
\begin{eqnarray*}\label{simu_mt2}
P(D=1|X_1,X_2,X_3)=\frac{\exp((\beta_1X_1+\beta_2X_2+\beta_3X_3)^3-2(\beta_1X_1+\beta_2X_2+\beta_3X_3))}{1+
\exp((\beta_1X_1+\beta_2X_2+\beta_3X_3)^3-2(\beta_1X_1+\beta_2X_2+\beta_3X_3))}.
\end{eqnarray*}
The parameter settings are the same as in Simulation I.

{\it Simulation III:} \quad
We consider the situation where the response model
and the treatment assignment model are no longer single-index models.
The response is generated by the following model:
\begin{eqnarray*}\label{simu_my2}
Y=\alpha D+(\gamma_1X_1+\gamma_2X_2+\gamma_3X_3)^2+X_2X_3+\epsilon.
\end{eqnarray*}
The treatment assignment model is
\begin{eqnarray*}\label{simu_mt2}
P(D=1|X_1,X_2,X_3)=\frac{\exp(\beta_1X_1+\beta_2X_2+\beta_3X_3+X_1^{2})}{1+
\exp(\beta_1X_1+\beta_2X_2+\beta_3X_3+X_1^{2})}.
\end{eqnarray*}
The parameter settings are the same as in Simulation I.

We generate $1000$ simulated data sets with the total sample size $n$ being  $300$ or $600$. The sample mean and sample standard deviation of $1000$ estimators
are given in the columns ``Mean'' and ``SD'', respectively.
The column ``ESD'' shows the estimated  standard deviation
and ``CI'' gives the nominal
$95\%$ confidence interval coverage rate using the estimated standard deviation.
We compare the proposed estimator $\widehat{\alpha}_{P}$ with two estimators:
$\widehat{\alpha}_{R}$, which is the estimator based on the  covariate adjustment
by propensity score (\citealp{vansteelandt2014regression,zou2016variance}); $\widehat{\alpha}_{L}$, which is the estimator based on the propensity score residuals \citep{lee2018simple}. The parameter $k$ related to  the proposed estimator $\widehat{\alpha}_{P}$ is chosen by mean square error. The simulation results are summarized
in Table 1.
$$\mbox{Table 1 about here}$$

From Table 1, we have the following observations. In Simulation I, the
scenario is  a relatively simple case.
 The three estimators
$\widehat{\alpha}_{R}$, $\widehat{\alpha}_{L}$ and $\widehat{\alpha}_{P}$
are all approximately unbiased.
The average of the standard error estimators of all
three estimators are close to
their respective sample standard deviation
and the confidence interval coverages are close to
 the nominal $95\%$ level.

 Simulation II considers more complex single index models.
 In this scenario,
the three estimators $\widehat{\alpha}_{R}$, $\widehat{\alpha}_{L}$ and $\widehat{\alpha}_{P}$ are all approximately unbiased.
 The average of the standard error estimators and
the confidence intervals of the above three estimators
 are close to the sample standard deviation and  attain coverage rate close to the nominal $95\%$ level, respectively.  The proposed estimator
 $\widehat{\alpha}_{P}$ is more efficient than
   $\widehat{\alpha}_{R}$ and $\widehat{\alpha}_{L}$.

When the response model
and the treatment assignment model are no longer single-index models in Simulation III, both $\widehat{\alpha}_{R}$ and   $\widehat{\alpha}_{L}$  are seriously  biased and consequently the $95\%$ confidence interval coverages
are much below the nominal level.
The proposed estimator $\widehat{\alpha}_{P}$ is approximately  unbiased.
The average of the standard error estimator of $\widehat{\alpha}_{P}$
is close to the sample standard deviation and the $95\%$
confidence interval coverage rate is close to
 the nominal level.

\section{An empirical application}
Low birth weight is one major determinant of infant morbidity and mortality and had been shown to be associated with prolonged negative effects on health and educational or labor market outcomes throughout life by many studies \citep{currie2011human}.
 \cite{kramer1987intrauterine} showed  that maternal smoking was the most important preventable negative cause of low birth weight. In this section, we use the dataset from Pennsylvania to
study the effect of maternal smoking on birth weight by the proposed method.

Pennsylvania dataset contains  observations of white mothers in Pennsylvania.
We focus on  observations of    non-Hispanic white mothers and the total sample size is $3,754$. The outcome of interest  $Y$ is infant birth weight, which is    measured in grams. The treatment assignment $D$ is a binary variable indicating
whether the mother smokes. The baseline covariates $X$ includes three
quantitative variables (mother's age,  mother's educational attainment and number of prenatal care visits) and four qualitative variables (indicator for alcohol consumption during pregnancy, indicator for the first baby, indicator for the first prenatal visit in the first trimester,  and indicator for
 a previous birth where the newborn died).

We consider three estimators $\widehat{\alpha}_{R}$, $\widehat{\alpha}_{L}$ and
  $\widehat{\alpha}_{P}$ that we considered in the simulation studies to study the effect of maternal smoking on birth weight.
   The estimators, standard errors, and
$p_{value}$ of $\widehat{\alpha}_{R}$, $\widehat{\alpha}_{L}$  and $\widehat{\alpha}_{P}$ are $(-271.657, 23.596, 0.000)$, $(-270.728, 23.267, 0.000)$ and $(-289.990, 22.656, 0.000)$, respectively. All the estimators confirm the maternal smoking has a significant impact on  low birth weight.

\section{Concluding remarks}

Correctly  identifying  treatment effects in observational studies is very difficult due to
the fact that the outcome model or the treatment assignment model must be correctly specified. To take  advantages of the semiparametric model,
the single-index models are used  to model the  relationship of the outcome and the treatment assignment between the baseline covariates.
We do not assume the link function is bounded and has bounded support
and use the sieve method to approximate the link function.
By the sieve method,
 the link function is regarded as a point in an
 infinitely dimensional function space, which allows us to estimate the link function
 and the index  parameter  simultaneously.
  Then, the average treatment effect can be estimated  by the simple linear regression. We establish the asymptotic properties of the proposed estimator.
 The finite-sample performance of the proposed estimator
 is evaluated through simulation studies. Simulation results show the proposed estimator
outperforms other commonly used competitor estimators.

 In this article, we consider  single-index models to estimate the treatment
 effect. It will be
 of interest to investigate using the
 Hilbert reproducible kernel space or deep learning methods
 to estimate the treatment effect in future work.

 \section*{Acknowledgements}
 This work is partly supported by
 the
Fundamental Research Funds for the Central Universities (Grant No: 31512111206)
and National Institutes of Health Grants P42ES031007 Super fund,
P30ES010126, and P01 CA142538 (for Zhou and Cai).

\bibliographystyle{apalike}

\bibliography{Semi_ATE}

\newpage
\begin{center}
\begin{table}
\caption{Simulation results in three  scenarios}
\begin{center}
\begin{tabular}{cccrcccccccc}
\hline \hline

&&& \multicolumn{4}{c}{$\alpha=0$} & \multicolumn{4}{c}{$\alpha=0.5$}
\\
\cline{4-7} \cline{9-12}
 Scenario & $n$ & Method     & Mean   &  SD  &  SE  & CI  & & Mean   &  SD  &  SE    & CI \\
\hline
I & 300 & $\widehat{\alpha}_R$ & -0.000 & 0.013  &0.013  &0.946  && 0.500  &0.013  &0.013  &0.946\\
&&        $\widehat{\alpha}_L$ & -0.000 & 0.013 &0.013   &0.939  && 0.500 &0.013   &0.013  &0.940\\
&&        $\widehat{\alpha}_P$ & -0.000 & 0.013 &0.021   &0.954  && 0.500 &0.013   &0.019  &0.954\\

&  600 &  $\widehat{\alpha}_R$ & -0.000  & 0.009  &0.009  &0.941  && 0.500  &0.009  &0.009  &0.941\\
&&        $\widehat{\alpha}_L$ & -0.000 & 0.009 &0.009   &0.932  && 0.500 &0.009   &0.009  &0.937\\
&&        $\widehat{\alpha}_P$ & -0.000 & 0.009 &0.015   &0.956  && 0.499 &0.009   &0.013  &0.950\vspace{2mm}\\

II & 300 & $\widehat{\alpha}_R$ & 0.001 & 0.193  &0.190  &0.956  && 0.501  &0.193  &0.190  &0.956\\
&&        $\widehat{\alpha}_L$ & -0.001 & 0.185 &0.188   &0.962  && 0.499 &0.185   &0.188  &0.962\\
&&        $\widehat{\alpha}_P$ & 0.002 & 0.144 &0.149   &0.954  && 0.500 &0.144   &0.148  &0.955\\

&  600 &  $\widehat{\alpha}_R$ & 0.002  & 0.144  &0.137  &0.946  && 0.502 &0.144  &0.137  &0.946\\
&&        $\widehat{\alpha}_L$ & 0.001 & 0.137 &0.135   &0.958  && 0.501 &0.137   &0.135  &0.958\\
&&        $\widehat{\alpha}_P$ & -0.001 & 0.103 &0.106   &0.955  && 0.497 &0.103   &0.107  &0.954\vspace{2mm}\\

III & 300 & $\widehat{\alpha}_R$ & 0.195 & 0.240  &0.234  &0.868  && 0.695  &0.240  &0.234  &0.868\\
&&        $\widehat{\alpha}_L$ & 0.147 & 0.241 &0.238   &0.899  && 0.647 &0.241   &0.238  &0.899\\
&&        $\widehat{\alpha}_P$ & 0.055 & 0.179 &0.174   &0.947  && 0.531 &0.174   &0.170  &0.953\\

&  600 &  $\widehat{\alpha}_R$ & 0.201 & 0.166  &0.167  &0.781  && 0.701  &0.166  &0.167  &0.781\\
&&        $\widehat{\alpha}_L$ & 0.142 & 0.172 &0.169   &0.861  && 0.642 &0.172   &0.169  &0.861\\
&&        $\widehat{\alpha}_P$ & 0.052 & 0.128 &0.125  &0.928  && 0.535 &0.127   &0.120  &0.929\\

 \hline\hline

\end{tabular}
\end{center}
\vspace{2mm}
Notation: $\widehat{\alpha}_R$ denotes the estimator based on propensity score regression, $\widehat{\alpha}_L$ denotes
the estimator based on propensity score residuals, $\widehat{\alpha}_R$ denotes our proposed estimator.
\end{table}
\end{center}

\end{document}